\documentstyle[11pt]{article}
\catcode`@=11
\def\fo{\hbox{{1}\kern-.25em\hbox{l}}}

\def\ch{\@startsection{section}{1}{\z@}{-3ex plus-1ex minus-.2ex}%
        {2ex plus.2ex}{\large\sc}}

\def\7#1#2{\mathop{\null#2}\limits^{#1}}        
\def\5#1#2{\mathop{\null#2}\limits_{#1}}        

\newcommand{\NP}[3]{{\sl Nucl. Phys.} (19#2)  {#3}}
\newcommand{\NPB}[3]{{\sl Nucl. Phys.} {\bf B#1} (19#2)  {#3}}
\newcommand{\JPA}[3]{{\sl J. Phys.} {\bf A#1} (19#2)  {#3}}

\newcommand{\PLB}[3]{{\sl Phys. Lett.} {\bf #1B} (19#2)  {#3}}

\newcommand{\PTP}[3]{{\sl Prog. Theor. Phys.} {\bf #1} (19#2)  {#3}}

\def\secteqno{\@addtoreset{equation}{section}%
\def\theequation{\thesection.\arabic{equation}}}
\def\endsecteqno{\def\theequation{\@ifundefined{chapter}%
{\arabic{equation}}{\thechapter.\arabic{equation}}}}
\newcounter{subequation}
\def\thesubequation{\alph{subequation}}
\def\sneqnarray{\stepcounter{equation}\let\@currentlabel=\theequation
\setcounter{subequation}{1}
\def\@eqnnum{{\rm (\theequation\thesubequation)}}
\global\@eqcnt\z@\tabskip\@centering\let\\=\@eqncr\let\@@eqncr=\@@sneqncr
$$\halign to \displaywidth\bgroup\@eqnsel\hskip\@centering
 $\displaystyle\tabskip\z@{##}$&\global\@eqcnt\@ne
 \hskip 2\arraycolsep \hfil${##}$\hfil
 &\global\@eqcnt\tw@ \hskip 2\arraycolsep $\displaystyle\tabskip\z@{##}$\hfil
  \tabskip\@centering&\llap{##}\tabskip\z@\cr}
\def\endsneqnarray{\@@sneqncr\egroup $$\global\@ignoretrue}
\def\@@sneqncr{\let\@tempa\relax
   \ifcase\@eqcnt \def\@tempa{& & &}\or \def\@tempa{& &}
   \else \def\@tempa{&}\fi
     \@tempa \if@eqnsw\@eqnnum\stepcounter{subequation}\fi
     \global\@eqnswtrue\global\@eqcnt\z@\cr}

\def\beq{\begin{equation}}
\def\eeq{\end{equation}}
\def\beqar{\begin{eqnarray}}
\def\eeqar{\end{eqnarray}}
\def\beqarn{\begin{eqnarray*}}
\def\eeqarn{\end{eqnarray*}}
\def\half{\frac{1}{2}}

\catcode`@=12

\secteqno

\title{ \vspace*{-1.0in}
\bf Field Redefinition Invariance  \\
and \\  ``Extra'' Terms}

\author{{\sc Karyn M. Apfeldorf  }$^\sharp \:\:$%
\thanks{apfel@utpapa.ph.utexas.edu}
{\sc , Carlos Ord\'o\~nez  }$^\flat\:\:$%
\thanks{cordonez@utaphy.ph.utexas.edu   $\:\:\:\:$ Current address:
University of Texas, Austin}
\\
\llap%
\small{\it Theory Group, Department of Physics  }$^{\sharp \:\: \flat}$ \\
\small{\it RLM\,5.208  University of Texas at Austin} \\
\small{\it Austin, TEXAS 78712} \\
\\
\llap%
\small{\it Department of Physics and Astronomy  }$^\flat$ \\
\small{\it Box 1807 Station B, Vanderbilt University } \\
\small{\it Nashville, TN 37235} \\
}
\date{}
\begin{document}
\maketitle
\thispagestyle{empty}
\begin{abstract}
\noindent
We investigate the issue of coordinate redefinition invariance
by carefully performing nonlinear transformations
in the discretized in quantum mechanical path integral.
By resorting to hamiltonian path integral methods, we provide the first
complete derivation of the extra term (beyond the usual jacobian
term) which arises in the action when a nonlinear transformation is made.
We comment on possible connections with the renormalization group,
by showing that these extra terms may emerge from a ``blocking'' procedure.
Finally, by performing field redefinitions before and after
dimensional reduction of a two dimensional field theory, we
derive an explicit form for an extra term appearing in a quantum field theory.
\end{abstract}
\vfill\hfill
\vbox{
\hfill July 1995 \null\par
\hfill UTTG-29-93}
\newpage

\section{Introduction}
Perhaps the most primitive of all properties of a physical theory is
invariance under redefinition of physical variables.
Indeed, experiments yield measurements and these numbers
relative to some scale have meaning, but the numbers in no
way imply a particular choice of variables.
One could go further to argue that in fact although
theoretical calculations using an effective physical theory
yield viable results, there is no proof that another framework
(perhaps not even quantum mechanics or quantum field theory)
could not yield similar or equivalent results.

Under coordinate or field redefinitions, physically meaningful quantities
must remain unchanged.  For example, one expects the poles of renormalized
propagators in quantum field theory to remain the same under a field
redefinition.
On the other hand, quantities related to choice of fields, such as
wave function renormalization factors, are physically insignificant
and may change.

In the case of quantum mechanics, it has been known for
decades~\cite{edwards} that
making nonlinear coordinate redefinitions generates ``extra''
(beyond the usual jacobian)
potential terms of ${\cal O}(\hbar^2)$.   From the hamiltonian point of
view
it is easy to understand why these so-called extra terms are
generated,  since upon quantization, the classical coordinates become
quantum operators,
and therefore there are nontrivial issues of operator ordering.
Much work has been published discussing the equivalence of the
hamiltonian and lagrangian approaches,~\footnote{
A given prescription for handling the discretized path integral
corresponds to a particular operator ordering of the associated
hamiltonian.  In particular, the midpoint prescription corresponds
to Weyl ordering, and in the coordinate representation supports
the interpretation of brownian motion.  For a nice review of the Feynman
path integral in quantum mechanics, see the work of Grosche~\cite{grosche}.}
and one would like also to understand the emergence of these terms from
the path integral point of view.
Essentially, this phenomenon of generating extra terms
is a manifestation of the stochastic nature
of the path integral, and may be studied from a discretization of the
quantum mechanical path integral.   The brownian nature of the paths
requires that one take care in evaluating the path integral (naive
substitution is insufficient), and one expects
extra terms, beyond the usual jacobian term, to be generated if the
coordinate transformation is nonlinear.

In contrast to quantum mechanics, the common practice in quantum field theory
is to ignore the possibility of any such extra terms, or to sweep them away
by appealing, in a hand-waving manner, to renormalization.
Motivated by the desire to explain this difference in practice, we
revisited the question of operator ordering in the quantum mechanical
path integral.
Gervais and Jevicki~\cite{gervais} investigated the
derivation of extra terms in quantum mechanics by path integral methods,
but we contend that there are deficiencies in their derivation.
Their derivation of the extra terms is
conspicuously missing some key steps in the
critical rewriting of the action.

In this paper, first we provide a solid path integral based
derivation of the extra terms in the case of quantum mechanics.
This is the first complete proof available and thus provides for an
extension to quantum field theory~\cite{us}.
Next, we indirectly investigate extra terms in quantum field theory by
connecting
to quantum mechanics.  More specifically, we examine what transpires
as one dimensionally reduces a $1+1$ dimensional field theory where
usually no extra terms are considered, to a quantum mechanical theory
where the existence of extra terms is well-established.
This example allows us to derive an explicit form of an extra term appearing in
a quantum field theory and serves as a starting point for a
more systematic study of extra terms in quantum field theory.
In a forth-coming article~\cite{us}, we argue that
in spite of the path integral's ubiquitous employment,
manipulations involving the path integral must be reevaluated,
and that interesting physics could be missed.

This paper is organized as follows.
In section 2, we revisit the issue of extra terms in quantum mechanics
by meticulously deriving the effects of a nonlinear point canonical
transformation in the discretized form of the quantum mechanical path
integral.
Unlike previous treatments~\cite{gervais} of this subject,
we give a complete treatment of all steps of the derivation.
We are thankful to B.~Sakita~\cite{sakitaPRI} for showing us an
important trick, using the hamiltonian form of the path integral,
to evaluate certain expectation values.
In section 3, we present a ``paradox'' where quantum field
theory meets quantum mechanics.  We consider
the dimensional reduction of a free, massless $1+1$ dimensional real scalar
quantum field theory to an effectively quantum mechanical theory,
and perform field redefinitions both before and after reduction.
This example allows us to confirm the existence of, and explicitly derive
the form of, the extra term generated in the quantum field theory action
upon a nonlinear change of field variables.
In section 4, we give conclusions and comment on further research questions.

\section{Nonlinear Coordinate Redefinitions in QM}

In this section, we
meticulously will derive the effects of a nonlinear point canonical
transformation in the discretized form of the quantum mechanical path
integral.

Although in some ways our derivation will resemble that of
Gervais and Jevicki~\cite{gervais}, it will be
different in at least one main respect.
Specifically, we will take a critical look at the
method of replacing terms involving difference variables
(corresponding to derivative interactions in the continuum limit)
which are generated in the action
after a coordinate transformation has been performed.

Consider the simple case of an $n$-dimensional quantum mechanical particle
with position $q^a(t).$  The path integral is
\beq
Z = \int \prod_{a=1}^n d q^a \: \:  e^{\frac{i}{\hbar} \int dt \left[ \half
\sum_{a=1}^n ({\dot q}^a)^2 - V_0(q) \right]} .
\eeq
In particular, we will assume that there are no
time derivatives in the potential $V_0(q)$.
Discretizing  the time interval $t_f - t_0$
into $N$ segments of length $\epsilon $ such that
$t_k = t_0 + \epsilon k $ and $t_f \equiv t_N,$
and using  the abbreviated notation
$q^a(k) = q^a(t_k) $, one arrives at
\beq
Z = \int \prod_{a=1}^n \prod_{k=1}^{N-1} d q^a (k)  \prod_{k=0}^{N-1}
e^{\frac{i}{\hbar} \left[ \frac{1}{2 \epsilon}
\sum_{a=1}^n ( q^a(k+1) - q^a(k)  )^2  - \epsilon V_0( q(k)) \right] }.
\eeq
In this expression, the product
$N \epsilon = t_f - t_0$ is held fixed, while the
limit $N \rightarrow \infty,$ $\epsilon \rightarrow 0$
is implied.
A key observation to make is that the one dimensional time
differential $dt$ carries a single power of
$\epsilon , $ so any terms of ${\cal O}(\epsilon^2)$
in the exponential will vanish in the small $\epsilon$ limit.
After a coordinate transformation,
the only ``extra'' terms that need be retained in the path
integral action are either ${\cal O}(\epsilon) $ or possibly divergent.

Under a coordinate redefinition $q^a(t) = F^a(Q(t))$
from the $q^a$ to the $Q^i$ ($i= 1,\ldots n$),
which translates to a point-by-point transformation on the
lattice $q^a(k) = F^a(Q(k))$, the path integral becomes
\beqar
Z &= & \int \prod_{i=1}^n \prod_{k=1}^{N-1} d Q^i (k)
\det \frac{\partial F^j(Q(k))}{\partial Q^\ell(k)}  \nonumber \\
& \times &  \prod_{k=0}^{N-1} e^{\frac{i}{\hbar} \left[ \frac{1}{2 \epsilon}
( F^a(Q(k+1)) - F^a(Q(k))  )^2  - \epsilon V_0( F(Q(k))) \right]} .
\eeqar
To rewrite the action in terms of the new variables,
with a ``canonical'' kinetic term for $Q$, we must expand
the function $F^a(Q(k))$ about some point of its argument.
The choice of a discretization procedure in the lagrangian path
integral corresponds
to a choice of an operator ordering prescription in the hamiltonian.
For concreteness, we choose the midpoint prescription, corresponding
to Weyl ordering; this is the only discretization scheme to preserve the
brownian motion interpretation of a quantum mechanical particle.
We will treat the jacobian determinant and the kinetic term
as follows. Define the midpoint and difference variables
\beqar
\bar{Q}^i(k) &\equiv& \half \left( Q^i(k+1) + Q^i(k) \right) \nonumber \\
\Delta Q^i(k) &\equiv&  Q^i(k+1) - Q^i(k) .
\label{eq:defmidpoint}
\eeqar
Invert these equations
\beqar
Q^i(k+1) &=& \bar Q^i(k) + \half \Delta Q^i(k)   \nonumber \\
Q^i(k) &=&   \bar Q^i(k) - \half \Delta Q^i(k)
\label{eq:invertdef}
\eeqar
and expand the $F^a(Q)$ about the midpoint variables, i.e.
\beqarn
F^a(Q(k)) &=& F^a(\bar Q(k)) - \half \Delta Q^i(k)
\frac{\partial F^a(Q(k))}{\partial Q^i(k)} |_{Q(k)=\bar Q(k)}  \\
& & +\frac{1}{2! 2^2} \Delta Q^i(k) \Delta Q^j(k)
\frac{\partial^2 F^a(Q(k))}{\partial Q^i(k) \partial Q^j(k)
} |_{Q(k)=\bar Q(k)} \\
& & - \frac{1}{3! 2^3} \Delta Q^i(k) \Delta Q^j(k) \Delta Q^\ell(k)
\frac{\partial^3 F^a(Q(k))}{\partial Q^i(k) \partial Q^j(k) \partial Q^\ell(k)
} |_{Q(k)=\bar Q(k)}
+ \cdots .
\eeqarn
Note that the transformed variable at a single lattice point $k$ gets
rewritten in terms of difference and midpoint variables at
both $k$ and $k+1$ lattice sites.
The original kinetic term in $q^a(k)$ becomes
\beqar
(\Delta q^a(k))^2 &=&  (q^a(k+1) - q^a(k))^2 = (F^a(Q(k+1)) -
F^a(Q(k)))^2  \nonumber \\
&=& \frac{\partial F^a(Q(k)) }{\partial Q^i(k)} |_{Q(k) = \bar Q(k)}
 \frac{\partial F^a(Q(k)) }{\partial Q^j(k)} |_{Q(k) = \bar Q(k)}
 \Delta Q^i(k) \Delta Q^j(k) \nonumber \\
& &  + \frac{1}{2 \cdot  3!}
 \frac{\partial F^a(Q(k)) }{\partial Q^i(k)} |_{Q(k)=\bar Q(k)}
\frac{\partial^3 F^a(Q(k)) }{\partial Q^j(k) \partial Q^\ell(k)
 \partial Q^m(k)}  |_{Q(k) = \bar Q(k)} \nonumber \\
& & \times \Delta Q^i(k) \Delta Q^j(k)  \Delta Q^\ell(k)  \Delta Q^m(k)
+ {\cal O} \left[ (\Delta Q)^5 \right] .
\eeqar
The metric residing in the canonical kinetic term for $Q^i(k)$ is
\beq
g_{ij}(\bar Q(k)) =
\frac{\partial F^a(Q(k)) }{\partial Q^i(k)} |_{Q(k) = \bar Q(k)}
\frac{\partial F^a(Q(k)) }{\partial Q^j(k)} |_{Q(k) = \bar Q(k)} .
\eeq
The metric must be a function {\it only} of the midpoint variables.
This point may be seen
especially clearly~\cite{sakita} from the hamiltonian point of view.
En route to treating the jacobian term, we evaluate
\beqar
& & \det \frac{\partial F^a(Q(k))}{\partial Q^\ell(k)}  \nonumber \\
&=&  [\det g_{\ell m} (\bar Q(k) ) ]^{\frac{1}{2}} \times
\exp \left[  -\frac{1}{4} \Delta Q^i(k) g^{\ell m}(\bar Q(k))
\frac{\partial \:}{\partial \bar Q^i(k)} g_{ m \ell}(\bar Q(k))
\right. \nonumber \\
& & \left. + \frac{1}{16}  \Delta Q^i(k)  \Delta Q^j(k)
\left(  g^{\ell m}(\bar Q(k)) \frac{\partial^2 \:}{\partial \bar
Q^i(k)\partial  \bar Q^j(k)} g_{m \ell}(\bar Q(k))  \right. \right.
\nonumber \\
& & \left.
\left. + \frac{\partial \:}{\partial \bar Q^i(k)}  g_{\ell m }(\bar Q(k))
\frac{\partial \:}{\partial \bar Q^j(k)} g_{m \ell}(\bar Q(k)) \right)
\right]
+ {\cal O} \left[ (\Delta Q)^3 \right] .
\eeqar
To maintain the symmetry of the action and treat the endpoints
correctly, we write for the jacobian
\beqar
& & \prod_{k=1}^{N-1}
\det \frac{\partial F^a(Q(k))}{\partial Q^\ell(k)} \\
&=& \prod_{k=1}^{N-1}   \left[
\det \frac{\partial F^a(Q(k))}{\partial Q^\ell(k)}  |_{Q(k) = \bar
Q(k) - \frac{1}{2} \Delta Q(k) } \times
\det \frac{\partial F^a(Q(k))}{\partial Q^\ell(k)}  |_{Q(k) = \bar
Q(k-1) + \frac{1}{2} \Delta Q(k-1) } \right]^{\frac{1}{2}}
\nonumber \\
&=&
\left[ \prod_{k=1}^{N-1}
\det \frac{\partial F^a(Q(k))}{\partial Q^\ell(k)}  |_{Q(k) = \bar
Q(k) - \frac{1}{2} \Delta Q(k) } \right]^{\frac{1}{2}}
\left[ \prod_{k=0}^{N-2}
\det \frac{\partial F^a(Q(k))}{\partial Q^\ell(k)}  |_{Q(k) = \bar
Q(k) + \frac{1}{2} \Delta Q(k) } \right]^{\frac{1}{2}}
\nonumber \\ &=&
\left[ \prod_{k=0}^{N-1}
\det \frac{\partial F^a(Q(k))}{\partial Q^\ell(k)}  |_{Q(k) = \bar
Q(k) - \frac{1}{2} \Delta Q(k) }
\det \frac{\partial F^a(Q(k))}{\partial Q^\ell(k)}  |_{Q(k) = \bar
Q(k) + \frac{1}{2} \Delta Q(k) } \right]^{\frac{1}{2}}
\nonumber \\  & & \times
\left[ \det \frac{\partial F^a(Q(k))}{\partial Q^\ell(k)}  |_{Q(k)^{=
\bar Q(0) - \frac{1}{2} \Delta Q(0) }_{= Q(0)}} \right]^{- \frac{1}{2}}
\left[ \det \frac{\partial F^a(Q(k))}{\partial Q^\ell(k)}  |_{Q(k)^{=
\bar Q(N-1) + \frac{1}{2} \Delta Q(N-1)}_{ = Q(N)}} \right]^{- \frac{1}{2}}
\nonumber \\  &=&
\prod_{k=0}^{N-1} \left[ \det g_{\ell m}(\bar Q(k)) \right]^{\frac{1}{2}}
\times \prod_{k=0}^{N-1} \exp \left[ \frac{1}{16} \Delta Q^i(k) \Delta
Q^j(k) {\cal G}_{ij}(\bar Q(k)) \right]
\nonumber \\ & &
\left[ \det g_{\ell m}(Q(0)) \right]^{-\frac{1}{4}}
\left[ \det g_{\ell m}(Q(N)) \right]^{-\frac{1}{4}}
\nonumber \eeqar
where
$$
{\cal G}_{ij}(\bar Q(k)) \equiv g^{\ell m}(\bar Q(k)) \frac{\partial^2
g_{m \ell}(\bar Q(k))}{\partial \bar Q^i(k)\partial  \bar Q^j(k)}
+ \frac{\partial g^{\ell m }(\bar Q(k)) }{\partial \bar Q^i(k)}
\frac{\partial g_{m \ell}(\bar Q(k)) }{\partial \bar Q^j(k)}.
$$

\noindent
Collecting the above results, one
sees clearly the canonical kinetic term and usual jacobian term
plus a series of additional ``extra'' terms.
Explicitly, one finds
\beqar
Z &=& \int \prod_{i=1}^n \prod_{k=1}^{N-1} d Q^i(k)
\prod_{k=0}^{N-1} \left[ \det g_{\ell m}(\bar Q(k))
\right]^{\frac{1}{2}}
\prod_{k=0}^{N-1}
e^{\frac{i}{\hbar} S[\bar Q(k),\Delta Q(k) ]}
\nonumber \\ & &
\times \left[ \det g_{\ell m}(Q(0)) \right]^{-\frac{1}{4}}
\left[ \det g_{\ell m}(Q(N)) \right]^{-\frac{1}{4}}
\eeqar
where
$$
S[\bar Q(k),\Delta Q(k) ] = \frac{1}{2 \epsilon} g_{ij}(\bar Q(k))
\Delta Q^i(k) \Delta Q^j(k)  - \epsilon V_0(F(\bar Q(k)))
- \epsilon V_{\rm extra}( \bar Q(k),\Delta Q(k))
$$
which implies the following ``extra'' terms beyond the usual jacobian
\beqar
V_{\rm extra}( \bar Q(k),\Delta Q(k))  &=&
\frac{i \hbar }{16 \epsilon} \Delta Q^i(k) \Delta Q^j(k)
{\cal G}_{ij}(\bar Q(k)) \\
& &  - \frac{1}{24 \epsilon^2}
\left[  \frac{\partial F^a(Q(k)) }{\partial Q^i(k)}
\frac{\partial^3 F^a(Q(k)) }{\partial Q^j(k) \partial Q^\ell(k)
 \partial Q^m(k)} \right]  |_{Q(k) = \bar Q(k)} \nonumber \\
&\times &
 \Delta Q^i(k) \Delta Q^j(k)  \Delta Q^\ell(k)  \Delta Q^m(k)
+ \cdots \nonumber .
\eeqar

In principle, one could stop at this point with a perfectly acceptable action.
However, this form of the action is not the most useful.  In
particular, the corresponding continuum version of the extra potential
in the action contains terms with multiple
derivatives of the coordinates, and the fact that the
series of terms in the difference variables truncates is not as clear.
To produce a superior form for the action, we will
``integrate out'' the difference variables
in the extra potential by a perturbation expansion and replace them with
functions of the midpoint variables.

Previous authors~\cite{gervais}, seeking to recast the
extra terms in the action into a non-derivative form,
have stated that one should employ the following integrals
(wherein $g = \det g_{ij} $)
\beqar
\label{eq:oldint}
\int \prod_{a=1}^n d x^a
e^{\frac{i}{\hbar} \frac{1}{2 \epsilon} g_{ij} x^i  x^j}  x^\ell x^m
&=& (2 \pi i \epsilon \hbar)^{n/2} g^{-\half}  \: (i \hbar \epsilon)
g^{\ell m}   \\
\int \prod_{a=1}^n d x^a
e^{\frac{i}{\hbar} \frac{1}{2 \epsilon} g_{ij} x^i  x^j}  x^\ell x^m x^p x^q
&=& (2 \pi i \epsilon \hbar)^{n/2} g^{-\half} (i \hbar \epsilon)^2
(g^{\ell m} g^{p q}+ g^{\ell p} g^{m q } + g^{\ell q} g^{m p})
\nonumber
\eeqar
to replace the difference variables by functions of the midpoint
variables.  The implication, of course, is that for each $k$,
$\Delta Q(k)$ is substituted in place of $x$.
Note that the above integrals may be simply derived from the
``normalization'' integral
\beq
\int \prod_{a=1}^n d x^a
e^{\frac{i}{\hbar} \frac{1}{2 \epsilon} g_{ij} x^i  x^j}
= (2 \pi i \epsilon \hbar)^{n/2} g^{-\half} .
\label{eq:normal}
\eeq
While Gervais and Jevicki~\cite{gervais} quote these integrals
in reference to this application,
and Sakita~\cite{sakita} gives a brief argument for use of these
integrals, any sort of proof or even a detailed accounting of the
procedure of replacing the difference variables with
functions of the midpoint variables is conspicuously lacking.
The critical issue is that the above prescription implies
that to integrate over the difference variables $\Delta Q(k)$ for each $k$,
we must have an integration factor $d \Delta Q(k)$ for each $k$.
However, this is not a straightforward procedure, as Sakita indeed
remarks in his book.

To understand the complication in applying the above formulas,
consider the base integral we must evaluate in order to derive the
integrals of powers of difference variables (dropping $V_0(\bar Q(k))$
since it has no consequence)
\beqar
Z_0 &\equiv& \int \prod_{i=1}^n \prod_{k=1}^{N-1} d Q^i(k)
\prod_{k=0}^{N-1} \left[ \det g_{\ell m}(\bar Q(k))
\right]^{\frac{1}{2}}
\prod_{k=0}^{N-1}
e^{\frac{i}{\hbar} \frac{1}{2 \epsilon} g_{ij}(\bar Q(k))
\Delta Q^i(k) \Delta Q^j(k) }
\nonumber \\ & &
\times \left[ \det g_{\ell m}(Q(0)) \right]^{-\frac{1}{4}}
\left[ \det g_{\ell m}(Q(N)) \right]^{-\frac{1}{4}} .
\label{eq:z0}
\eeqar
The problem is that this integral is not analogous to the
``normalization'' integral listed in the left hand side of
equation~\ref{eq:normal}.
To correctly apply those integrals above, we must rewrite the
$d Q(k)$ differentials in terms of the $d \Delta Q(k)$ differentials.
Applying the expressions in equation~\ref{eq:invertdef}
for $Q(k)$ and $Q(k+1)$ in terms of the midpoint and difference variables
to the differentials of
two neighboring coordinates $Q(k)$ and $Q(k+1)$, we find
\beq
dQ(k) dQ(k+1) = d \bar Q(k) \: d \Delta Q(k).
\eeq
Note that if we just apply this along the chain of differentials,
we get
\beq
dQ(1) dQ(2) dQ(3) dQ(4)  \ldots = d \bar Q(1) \: d \Delta Q(1) d \bar
Q(3) \: d \Delta Q(3)  \ldots ,
\eeq
or in other terms, taking an even number of lattice points, we find
\beq
\prod_{i=1}^{2 M}  dQ(i) = \prod_{i=1}^M d \bar Q(2i-1) \: d \Delta
Q(2i-1) .
\eeq
Note that we get differentials of midpoint and difference variables labeled
with every other point.  In particular, we do not get a differential
of a difference variable at each point.  We believe that this
feature, which is reminiscent of the the ``blocking'' of a lattice,
may be the root of
a connection between field redefinitions and the renormalization
group.
This matter is currently being investigated by the authors~\cite{us}.

Instead of working with the lagrangian form of the path integral,
we will turn to the hamiltonian form of the path integral
to evaluate the necessary integrals.
We are grateful to B. Sakita~\cite{sakitaPRI} for pointing out
to us the utility of the phase space method in this case.
Consider the phase space version of the path integral given in
equation~\ref{eq:z0}, with action given by
\beq
S = \sum_{k=0}^{N-1} \left[ P_i(k) \Delta Q^i(k)
- \frac{\epsilon}{2} g^{ij}(\bar Q(k)) P_i(k) P_j(k) \right] .
\eeq
Dropping irrelevant normalization factors and using the abbreviation
$[d Q ][d P] = \prod_{k=1}^{N-1} d Q(k) \prod_{k=0}^{N-1} d P(k) $, we
use the following relation
\beqar
& & \int [d Q ][d P]
e^{- \frac{i \epsilon }{2 \hbar} \sum_{k=0}^{N-1}
g^{ij}(\bar Q(k)) P_i(k) P_j(k) }
\frac{ \hbar \partial \:\:\:\:}{i \partial P_\ell(k) }
\frac{ \hbar \partial \:\:\:\:}{i \partial P_m(k) }
e^{ \frac{i}{\hbar} \sum_{k=0}^{N-1} P_n(k) \Delta Q^n(k) }
\nonumber \\ &=&
\int [d Q ][d P]
e^{ \frac{i}{\hbar} \sum_{k=0}^{N-1} P_n(k) \Delta Q^n(k) }
\frac{ \hbar \partial \:\:\:\:}{i \partial P_\ell(k) }
\frac{ \hbar \partial \:\:\:\:}{i \partial P_m(k) }
e^{- \frac{i \epsilon }{2 \hbar} \sum_{k=0}^{N-1} g^{ij}(\bar Q(k))
P_i(k) P_j(k) } \nonumber
\eeqar
to derive the exact relation
\beq
\langle \Delta Q^\ell (k) \Delta Q^m (k) \rangle_{\rm {Phase}}
= i \epsilon \hbar \langle g^{\ell m}(\bar Q(k)) \rangle_{\rm {Phase}}
+ \epsilon^2 \langle g^{\ell i}(\bar Q(k)) g^{m j }(\bar Q(k))
P_i (k) P_j (k)  \rangle_{\rm {Phase}}  .
\label{eq:exact}
\eeq
Our plan is to prove that the second term on the right-hand side is
${\cal O}(\epsilon^2),$ i.e. the term inside the brackets is of
${\cal O}(1)$.  If this is true, then the second term may be dropped,
since only terms of ${\cal O}(\epsilon)$ survive in the action.
Toward this end, we will employ phase space path integral
perturbation expansion techniques.  Following loosely the
notation of Sakita's book~\cite{sakita}, the generating functional is
\beq
Z[J,K] = \lim_{\begin{array}{l} t_f \rightarrow  \infty \\
    t_i \rightarrow -\infty \end{array}}
\int \int dQ_f dQ_i
\psi_0^*(Q_f,t_f) {\cal K}(Q_f,t_f,Q_i,t_i) \psi_0(Q_i,t_i)
\eeq
where ${\cal K}(Q_f,t_f,Q_i,t_i)$ is the Feynman kernel and
$\psi_0(Q,t)$ is the
asymptotic ground state wave function of the free hamiltonian.
Explicitly, the Feynman kernel is
\beq
{\cal K} = \int \prod_{k=1}^{N-1} d Q(k) \prod_{k=0}^{N-1} \frac{d P(k)}{2 \pi}
\exp{i S[J,K]}
\eeq
where $S[J,K]$ is the phase space action with sources $J$ and $K$ for
$Q$ and $P$ respectively
$$
S[J,K] = \left[ \sum_{n=0}^{N-1} \epsilon \left( P(n) \dot Q(n) - H_n \right)
+ \epsilon J(n) Q(n) + \epsilon K(n) P(n)  \right].
$$
Writing the hamiltonian as
$$
H_n = H_0(n) + H_1 \left( P(n), \frac{Q(n)+Q(n-1)}{2} \right),
$$
and taking the free hamiltonian to be
$$H_0 = \frac{P^2 + \omega^2  Q^2}{2}$$
the asymptotic ground state wave function may be calculated to be
$$
{\psi_0} = (\frac{\omega}{\pi})^\frac{1}{4} e^{-\half \omega Q^2 -
\frac{i}{2} \omega t}.
$$
To develop a perturbation expansion, we write
\beq
Z[J,K] = e^{\int_{-\infty}^\infty  dt H_1[\frac{1}{i} \frac{\partial
\:}{\partial K}, \frac{1}{i} \frac{\partial \:}{\partial J}]}
Z_0[J,K],
\eeq
and subsequently calculate the generating functional $Z_0[J,K]$ to be
\beqar
Z_0[J,K] &=& \lim_{
t_f \rightarrow \infty , t_i \rightarrow -\infty
}
(\frac{\eta}{2})^{\frac{N}{2}} \left( i e^{i \omega (t_f-t_i)-N
\alpha} \right)^\half \\
&\times& e^{\left[ -\frac{\epsilon^2}{2} \sum_{n,m=1}^N
[ J(n) - \dot K(n+1)] \Delta_F(n,m) [ J(m) - \dot K(m+1)]
+\frac{i}{2} \epsilon \sum_{n=1}^N K^2(n) \right]} \nonumber
\eeqar
($ \dot K(n+1) = \frac{ K(n+1) - K(n)}{ \epsilon} $) where
$$
\eta = \frac{8 \epsilon}{4 + \omega^2 t^2} = \frac{2}{\omega} \sin
\alpha $$
and
$$\tan \frac{\alpha}{2} = \frac{\epsilon \omega}{2}.$$
The lattice propagator is
\beq
\Delta_F(n,m) = \frac{1}{2 \omega} [ \theta(n-m-1) e^{-i(n-m) \alpha}
+ \delta_{n,m} +  \theta(m-n-1) e^{-i(m-n) \alpha}]
\eeq
where $\theta_\ell $ is the discrete version of the theta function,
i.e. $\theta_\ell = 1$ if $\ell \geq 0$ and $\theta_\ell = 0$ if $\ell < 0$.
Some useful relations for the following discussion are
\beqar
\Delta_F(m,m) &=& \frac{1}{2 \omega} \:\:\:\:\:\: \:\:\:\:\:\:\:
 \forall \:\: m \nonumber \\
\Delta_F(m,m-1) = \Delta_F(m-1,m)
 &=& \frac{1}{2 \omega} e^{-i \alpha} \:\:\:\:\:\: \forall \:\: m \nonumber .
\eeqar
Using the above, one may evaluate the phase space propagators
\beq
\langle Q(n) Q(m) \rangle_{\rm Phase} =  \frac{1}{(i \epsilon)^2}
\frac{\delta^2 Z_0[J,K]}{\delta J(n) \delta J(m)} |_{J=0,K=0}  =
\Delta_F(n,m) = \Delta_F(m,n)
\eeq
\beq
\langle Q(n) P(m) \rangle_{\rm Phase} =  \frac{1}{(i \epsilon)^2}
\frac{\delta^2 Z_0[J,K]}{\delta J(n) \delta K(m)} |_{J=0,K=0}
=    \frac{1}{\epsilon} \left[  \Delta_F(n,m) - \Delta_F(n,m-1)
\right]
\eeq
and
\beqar
\langle P(n) P(m) \rangle_{\rm Phase} &=&  \frac{1}{(i \epsilon)^2}
\frac{\delta^2 Z_0[J,K]}{\delta K(n) \delta K(m)} |_{J=0,K=0} \nonumber  \\
&=& \frac{1}{i \epsilon} \delta_{nm}  +
\frac{1}{\epsilon^2} \left[  \Delta_F(n,m) +  \Delta_F(n-1,m-1)
\right.   \nonumber  \\
& &  \left.  - \Delta_F(n-1,m) - \Delta_F(n,m-1) \right].
\eeqar
Determining the order in $\epsilon$ of the second term of
equation~\ref{eq:exact} requires that we calculate
$\langle \bar Q(k) \bar Q(k) \rangle_{\rm Phase}$,
$\langle \bar Q(k) P(k) \rangle_{\rm Phase}$, and
$\langle P(k) P(k) \rangle_{\rm Phase}$.
Firstly, note that a consequence of using the midpoint coordinate
$\bar Q(m)$, as opposed to $Q(m)$, gives
\beqar
& & \langle \bar Q(m) P(m) \rangle_{\rm Phase}
 = \half \langle Q(m) P(m) \rangle_{\rm Phase} +
\half \langle Q(m-1) P(m) \rangle_{\rm Phase}  \nonumber \\
&=& \frac{1}{2 \epsilon} \left[  \Delta_F(m,m) -  \Delta_F(m,m-1)
      + \Delta_F(m-1,m) - \Delta_F(m-1,m-1) \right]  \nonumber \\
&=& \frac{1}{2 \epsilon} \left[  (\Delta_F(m,m) -  \Delta_F(m-1,m-1))
      + (\Delta_F(m-1,m) - \Delta_F(m,m-1)) \right]  \nonumber \\
&=& 0 .
\eeqar
Secondly, since the two-point function of two $Q(m)$'s is
independent of $\epsilon$, so must be the two-point function
of two  $\bar Q(m)$'s since the $\bar Q(m)$'s  are linear combinations
of the $Q(m)$'s.
Finally, one may calculate the two-point function of two $P(m)$'s
taken at the same point
\beqar
\langle P(m) P(m) \rangle_{\rm Phase}
 &=& \frac{-i}{\epsilon} + \frac{1}{\epsilon^2} \left[
\Delta(m,m) + \Delta(m-1,m-1) - \Delta(m-1,m) - \Delta(m,m-1) \right]
\nonumber \\
&=& \frac{-i}{\epsilon} + \frac{1}{\epsilon^2 \omega} (1 - e^{-i \alpha})
\nonumber \\
&=& -\frac{\omega}{2} - \frac{i}{6} \epsilon \omega^2 + \cdots .
\eeqar
Together with Wick's theorem, these expectation integrals show that
the second term on the right hand side of equation~\ref{eq:exact} is order
${\cal O}(\epsilon^2)$ and thus is of no consequence.
Since the second term may be neglected, and the
remaining terms do not contain averages over $P(m)$'s, we may implicitly
integrate over the $P(m)$'s in the phase space expectation values, to
arrive at a lagrangian formula valid in the same small $\epsilon$ limit
\beq
\langle \Delta Q^\ell (k) \Delta Q^m (k) \rangle
= i \epsilon \hbar \langle g^{\ell m}(\bar Q(k)) \rangle.
\eeq
In the same way, one may prove that
\beqar
& & \langle \Delta Q^i (k) \Delta Q^j (k) \Delta Q^\ell (k) \Delta Q^m (k)
\rangle  \\
&=&   (i \epsilon \hbar)^2  \langle g^{ij}(\bar Q(k))  g^{\ell m}(\bar Q(k))
+ g^{i \ell}(\bar Q(k))  g^{j m}(\bar Q(k))
+ g^{i m}(\bar Q(k))  g^{j \ell}(\bar Q(k))
\rangle. \nonumber
\eeqar
Finally, by using the preceeding two relations to replace the difference
variables with functions of the midpoint variables, we arrive at a
contribution (at midpoint site $\bar Q_i$)
\beq
S_{\rm extra}[\bar Q_i] =  - \hbar^2 \epsilon
\sum_{k,\ell, p, n \in Z} \frac{1}{8}
g^{\ell n}( \bar Q_i) \Gamma^k_{\ell p} (\bar Q_i) \Gamma^p_{nk} (\bar Q_i)
\eeq
which is written compactly here in terms of the metric and connection of
the coordinate transformation.
In the continuum limit, this amounts to an extra potential term
\beq
S_{\rm extra}[Q] =  - \hbar^2 \int dt
\sum_{k,\ell, p, n \in Z} \frac{1}{8}
g^{\ell n}(Q) \Gamma^k_{\ell p} (Q) \Gamma^p_{nk} (Q)
\label{eq:gg} .
\eeq
This contribution only occurs for
nonlinear transformations, since while the metric involves only one
derivative of the transformation, the connection involves two derivatives.
This form for the extra term agrees with the previously obtained result
of Gervais and Jevicki~\cite{gervais}.

The important point of this analysis in the quantum mechanical case
lies not so much in this final
form for the extra term generated by a nonlinear coordinate
redefinition, but in our derivation and its consequences.
We have provided a solid path integral proof for the extra term,
and have supplied conspicuously missing steps in the derivation.
In the process, we have uncovered a possible connection between
field redefinitions and renormalization group transformations,
by way of ``blocking'' the lattice.
Perhaps most importantly, the successful path integral approach
will allow a similar analysis in the quantum field theory case.

In the following section, we indirectly explore nonlinear field
redefinitions in quantum field theory by
bridging the gap between quantum mechanics
and quantum field theory via dimensional reduction.  This serves as
a concrete example for the generation of extra terms in quantum field
theory and as motivation for our future paper in which we give a
direct, complete analysis using discretization of the quantum field
theory path integral~\cite{us}.

\section{Kaluza-Klein ``Paradox''}

In contrast to the case of quantum mechanics,
the standard lore in quantum field theory dictates
that under a field redefinition, the action changes by direct
substitution of the change of variables in the action together with
inclusion of the jacobian determinant of the transformation.
Furthermore, when dimensional regularization is employed, the
jacobian determinant does not contribute since upon exponentiation
the formally infinite spacetime delta function $\delta^{(d)} (0)$ generated by
the trace is set equal to zero.  Similarly, there is also a standard
argument that any extra terms
generated in the path integral, since they are manifestations of
operator ordering (i.e. from $ [ \pi(x) ,\phi (x) ]  =
-i \hbar \delta^{(d-1)}(0)$), would involve delta functions at zero argument
and therefore would vanish by dimensional regularization.
Even if the validity of dimensional regularization is not questioned, a
solid justification for setting infinite quantities equal to zero is lacking.
Certainly from the lattice point of view, the jacobian term, as well as any
extra term, is very real and does not vanish since the spacetime delta
function at zero argument
is just a power of the inverse lattice spacing i.e.,
$\delta^{(d)}(0) \sim  a^{-d}.$
Dimensional regularization does in fact fail in cases where the action
has some feature depending on the dimensionality of spacetime.

An effective laboratory in which to learn about nonlinear point
canonical transformations in quantum field theory is that of a
1+1 dimensional real scalar field on a Minkowskian cylinder.
By integrating out the angular coordinate one arrives at an
effectively quantum mechanical problem, and then the
well-established results for nonlinear point canonical transformations
in quantum mechanics may be employed.
On the other hand, one may use the standard lore to make a
nonlinear field redefinition directly in the 1+1 dimensional
quantum field theory, and then afterwards integrate out the
angular coordinate to arrive again to a quantum mechanics action.
One expects that the two approaches should
give the same result, so if it is true that no new terms appear
in the latter scenario when one performs a field redefinition in the
quantum field theory, one must be able to explain
what happens to the extra terms that arise along the way in
the former scenario.
In the following we consider a real massless scalar field $\phi(x,t)$
in a flat $1+1$ Minkowskian spacetime with a periodic spatial coordinate
\beq
S = -\half \int d^2 x \:\: \eta^{\mu \nu} \partial_\mu \phi \partial_\nu \phi
\eeq
where
$\eta_{tt} = 1,$  $\eta_{x x} =-1$ and $x \cong x + 2 \pi R . $

\subsection{Method 1}

Using an angular coordinate $\theta = x/R,$ the action is
\beq
S = R  \int d\theta dt \left[ \half \dot{\phi}^2
- \frac{1}{2 R^2} (\partial_\theta \phi)^2 \right] .
\eeq
One may convert this quantum field theory into an effectively
quantum mechanical problem by expanding the scalar field in a
Kaluza-Klein-like decomposition
\beq
\phi(\theta,t) = \sum_{p \in Z} a^{(p)}(t) e^{i p \theta}
\eeq
and integrating out the $\theta$ dependence in the action.
The resulting action is that of a sum of an infinite number of
quantum mechanical complex oscillators $a^{(m)}(t)$ with frequencies
$\omega_m = \frac{m}{R}$
dependent on mode number $m$ and radius $R$ of the internal dimension.
Explicitly,
\beq
S_0[a]  = \sum_{p \in Z} \int dt \left[ \sum_{m \in Z}
\half g_{mp}[a] \dot{a}^{(m)} \dot{a}^{(p)}
- \half \frac{p^2}{R^2} g_{mp}[a] a^{(m)} a^{(p)} \right]
\label{eq:free}
\eeq
where the
metric
$$g_{mp}[a] = 2 \pi R \delta_{m,-p} = 2 \pi R \delta_{m+p=0}$$
is an infinite-dimensional unit matrix multiplied by $2 \pi R$.
We may consider a ``cutoff'' version of this theory, wherein
only modes with $|n| \leq N_{\rm max}$ are included.  In this case,
$g_{mp}[a]$ is an ordinary $2N_{\rm max} +1$ by $2N_{\rm max} +1$
metric defining a Euclidean space of oscillators $a(t).$

Let us perform a change of variables in this quantum mechanical model.
In particular, we wish to make the change of variables corresponding to
\beq
\phi = F[\varphi] \equiv  \varphi + \alpha \varphi^N .
\label{eq:Ntran}
\eeq
We expand the new field $\varphi (\theta , t)$ in modes
\beq
\varphi(\theta,t) = \sum_{p \in Z} b^{(p)}(t) e^{i p \theta} ,
\eeq
and define the following useful objects
\beqar
E^{(\ell)}_L [b]   &\equiv& \sum_{p_1 \in Z} \ldots \sum_{p_L \in Z}
b^{(p_1)} \ldots b^{(p_L)} \delta^{\ell - \sum_{i=1}^L p_i =0} \\
B_L [ b ] &\equiv&  E^{(0)}_L [ b ] = \int \frac{d\theta}{2 \pi}
\varphi(\theta, t)^L ,
\label{eq:defB}
\eeqar
where $\delta^{\ell=0}$ is one when $\ell=0$ and is zero if $\ell \neq 0.$
Note $E^{(\ell)}_L [b] $ involves the product of $L$ $b$'s whose mode
numbers add up to $\ell,$ and $B_L [b] $ involves $L$ $b$'s whose
overall sum of mode numbers vanishes.
A number of identities, the simplest being
\beqar
\sum_{p \in Z} E^{(\ell + p)}_M [b] E^{(-p+m)}_L [b] &=&
E^{(\ell+m)}_{M+L} [b]
\label{eq:ident1}  \\
\frac{\partial \:\:}{\partial b^{(m)} } E^{(\ell)}_M [b] &=&
M E^{(\ell-m)}_{M-1} [b]
\eeqar
greatly facilitate subsequent calculations.  Appendix A contains
useful identities and their proofs.
The transformation of the modes corresponding to the field redefinition
above in equation~\ref{eq:Ntran} is
\beq
a^{(p)} \equiv f^{(p)} [b]  = b^{(p)} + \alpha E^{(p)}_N [b] .
\label{eq:atob}
\eeq
To implement this change of coordinates in the quantum mechanical
path integral, we must include not only the naive substitution
of change of variables in the action and the jacobian determinant, but also
the extra term due to the stochastic nature of the path integral.
Specifically, after
exponentiating the usual jacobian factor, the action becomes
\beq
S[b]=S_0[b] + S_{\rm jacobian}[b] + S_{\rm extra} [b]
\eeq
where $S_0 [b ] $  is the action obtained by direct substitution into
the original free action given in equation~\ref{eq:free},
and where, as derived in equation~\ref{eq:gg},
the extra term of ${\cal O}(\hbar^2)$ is
\beq
S_{\rm extra} [b] = - \hbar^2 \int dt \sum_{k,\ell, p, n \in Z}
\frac{1}{8} g^{\ell n} [b] \Gamma^k_{\ell p}[b]  \Gamma^p_{nk}[b] .
\eeq
It is possible to compute all three contributions to the
action $S[b]$ exactly to all orders in the parameter $\alpha$.
The metric in the new coordinates is given by
\beq
g_{mn}[b] =  2 \pi R \left(
\delta_{m+n=0} + 2 \alpha N E^{(-m-n)}_{N-1}[b] + N^2 \alpha^2
E^{(-m-n)}_{2N-2}[b] \right) ,
\eeq
and the inverse metric, which may be obtained via a recursion equation
(see appendix B), is given by
\beq
g^{km}[b] = \frac{1}{2 \pi R} \sum_{j=0}^\infty (-\alpha N)^j (j+1)
E^{(k+m)}_{j(N-1)}[b].
\eeq
In terms of this metric, direct substitution of the nonlinear change
of variables into the original action yields
\beq
S_0[b]  = \sum_{p \in Z} \int dt \left[
  \half \sum_{m \in Z} g_{mp}[b] \dot{b}^{(m)} \dot{b}^{(p)}
- \half \frac{p^2}{R^2} 2\pi R f^{(p)}[b] f^{(-p)}[b] \right] ,
\label{eq:m10}
\eeq
where $f^{(p)}[b]$ was given in equation~\ref{eq:atob}.
The jacobian factor may be evaluated (see appendix B) to give a
contribution to the action of
\beqar
S_{\rm jacobian}[b]
& =& -\frac{ i \hbar}{2}  {\rm Tr ln} g_{mn}[b] \delta (t -t^\prime )
\nonumber \\
& =&   i \hbar  \delta(0)
\left[ \sum_{k \in Z} 1 \right]
\sum_{j=1}^{\infty} \frac{(-\alpha N)^j}{j}
\int dt B_{j(N-1)}  ,
\label{eq:m1j}
\eeqar
where we have dropped an infinite constant
$- i\hbar \delta(0) {\rm ln }[2 \pi R] [\sum_{k \in Z} 1 ] /2 .$
Finally, the extra term arising from the stochastic
nature of the path integral may be evaluated (see appendix B) to give
a contribution to the action of
\beqar
S_{\rm extra}[b] &=& -\frac{\hbar^2}{8}
\frac{(N-1)^2}{2 \pi R} \left[ \sum_{k \in Z} 1 \right]^2
\sum_{X=0}^\infty (-\alpha N)^{X+2}
\left( \begin{array}{c} X+3 \\ 3 \end{array} \right)  \nonumber \\
& & \times \int dt \: \:   B_{X(N-1)+2(N-2)}[b]  .
\label{eq:m1e}
\eeqar
Note the presence of the infinite sum over
modes in equation~\ref{eq:m1j} and the same sum squared in
equation~\ref{eq:m1e}.

\subsection{Method 2}
Alternately, we may choose to perform the field redefinition in the original
quantum field theory, and then afterwards expand in the Kaluza Klein modes.
Upon direct substitution, the field redefinition of equation~\ref{eq:Ntran}
introduces derivative interaction terms into the action
\beq
S_0[\varphi] = -\half \int d^2 x  \; \eta^{\mu \nu}
(1+ 2 N \alpha \varphi^{N-1} + N^2 \alpha^2 \varphi^{2(N-1)} )
\partial_\mu \varphi (x) \partial_\nu \varphi (x) .
\eeq
The exponentiated jacobian determinant gives a contribution
to the action
\beq
S_{\rm jacobian}[\varphi] = - i \hbar \delta^{(2)}(0) \int d^2 x \:
{\rm ln}[1 + N \alpha \varphi^{N-1}] .
\eeq
Assuming the conventional arguments hold would imply that
the total action is given by $S_0[\varphi] + S_{\rm jacobian}[\varphi].$
We now convert this action to an effectively quantum mechanical one by
integrating out the angular coordinate.
The terms in the action for the $b$ modes obtained from method
2 will be written with tildes to distinguish them from those of
method 1.
Upon integrating out the $\theta$ dependence, $S_0[\varphi]$ becomes
\beq
\tilde S_0[b] =  \int dt \sum_{m \in Z} \sum_{p \in Z}
\left[ \half  g_{mp}[b]  \dot{b}^{(m)} \dot{b}^{(p)}
+ \frac{m p}{2 R^2} g_{mp}[b]  b^{(m)} b^{(p)} \right]
\label{eq:m20}
\eeq
where $g^{(b)}_{mp}$ is exactly the metric appearing in the
previous section.
Similarly, expanding the logarithm and using equation~\ref{eq:defB},
one obtains the expression for the jacobian
\beq
\tilde S_{\rm jacobian}[b] =  2 \pi R i \hbar \delta^{(2)}(0)
\sum_{j=1}^\infty  \frac{(-\alpha N )^j}{j}  \int dt  B_{j(N-1)} .
\label{eq:m2j}
\eeq

\subsection{Comparison of Methods}

If the standard lore holds true, one would expect to find that
the two actions for the effective quantum mechanics theory
are equivalent, i.e.
\beq
S_0[b] + S_{\rm jacobian}[b] + S_{\rm extra} [b] = \tilde S_0[b] +
\tilde S_{\rm jacobian}[b]  .
\eeq
Firstly, we note that using identity~\ref{eq:p2ident} from
appendix A allows one to prove the equality of the kinetic terms, i.e.
$S_0[b] = \tilde S_0[b]$ (see equations~\ref{eq:m10} and~\ref{eq:m20} ) .

Secondly, comparison of the jacobian factors
(equations~\ref{eq:m1j} and~\ref{eq:m2j}) indicates that the following
correspondence must hold
\beq
 \sum_{k \in Z} 1  =  2 \pi R \: \: \delta (0)
\label{eq:sumtodelta}
\eeq
where the delta function at zero argument on the right hand side of this
equation is a delta function in the coordinate $x$, i.e.
$\delta(0) = \5{x \rightarrow 0}{\rm Lim } \delta(x).$  If we express
the delta function in terms of the angular coordinate $\theta = x/R$,
the $R$ dependence in the above equation drops out.
With this identification, we find $ S_{\rm jacobian}[b]
= \tilde S_{\rm jacobian}[b]$ up to the infinite constant that we dropped.

Finally, we turn our attention to the most interesting term, namely
$S_{\rm extra}[\varphi].$
In the second method where we performed the nonlinear field
redefinition in the quantum field theory using the standard arguments,
we did not pick up any such term.
Taking the pragmatic point of view that this term can not
be rationalized to vanish, we ask the following question
``could this term arise as an extra term generated upon nonlinear
field redefinition in the original quantum field theory?''
Remarkably, it is possible (see appendix B) to perform the resummation
of the infinite series in equation~\ref{eq:m1e}.  The resulting term is
\beq
S_{\rm extra}[\varphi] = -\frac{\hbar^2}{8}
\frac{\left[ \sum_{k \in Z} 1 \right]^2}{(2 \pi R)^2} \alpha^2 N^2 (N-1)^2
\int d^2 x \frac{\varphi^{2N-4}}{(1+\alpha N\varphi^{N-1})^4} .
\label{eq:VE}
\eeq
We still must contend with the peculiar infinite summations to
relate this potential to the two dimensional field theory.
Using the result of the comparison of the jacobian terms, i.e.
equation~\ref{eq:sumtodelta}, we conclude that the extra term
becomes
\beq
S_{\rm extra}[\varphi] = -\frac{\hbar^2}{8}  (\delta(0))^2
\alpha^2 N^2 (N-1)^2
\int d^2 x \frac{\varphi^{2N-4}}{(1+\alpha N\varphi^{N-1})^4} .
\label{eq:kkextra}
\eeq
Interestingly, note that this term diverges as the
square of the $\delta (0) $ in the spatial coordinate,
and {\it not} like the two-dimensional delta function at zero argument.
This feature (which persists in $d$-dimensions quantum field theory where
an extra term containing $(\delta^{(d-1)} (0))^2$ appears)
occurs for an important reason which becomes evident when one
evaluates Feynman diagrams.

We could have elected to present the material in the intermediate step
of Method 2 of this exercise in a different, more complete manner.
That is, after the field redefinition, we should add to the action an
infinite series of counterterms with unspecified coefficients
$c_\ell,$ i.e.
\beqarn
S[\varphi] &=& -\half \int d^2 x  \; \eta^{\mu \nu}
(1+ 2 N \alpha \varphi^{N-1} + N^2 \alpha^2 \varphi^{2(N-1)} )
\partial_\mu \varphi (x) \partial_\nu \varphi (x)  \\
& & - i \hbar \delta^{(2)}(0) \int d^2 x \:
{\rm ln}[1 + N \alpha \varphi^{N-1}]
+ \int d^2x \: \sum_{\ell=1}^\infty c_\ell \varphi^\ell (\varphi) .
\eeqarn
With this action, one would have to determine the infinite series of
unknown coefficients, the $c_\ell$'s, by calculating physically significant
quantities and matching with the original free theory.
On the other hand, our exercise has enabled us to pin down the precise
counterterm without having to resort to evaluating Feynman diagrams, i.e.
the series in $c_\ell$ is replaced by the simple expression in
equation~\ref{eq:kkextra}.

To conclude this section, we have given concrete evidence for a
single~\footnote{The fact that only a single extra term appears in
this example is related to the particular spacetime topology in the problem.
Our future works will treat quantum field theories in
generic spacetimes.}
extra term to be generated upon making a nonlinear field redefinition
in this $1+1$ dimensional quantum field theory
on a Minkowskian cylinder.
With this result as motivation, we are poised to
we tackle this subject directly in a subsequent paper~\cite{us}, by
discretizing the quantum field theory path integral.

\section{Conclusion}

We have provided the first solid path integral based
derivation of the extra potential term, beyond the usual jacobian term,
which arises when one makes nonlinear coordinate redefinitions
in quantum mechanics.
By carefully performing nonlinear transformations
in the discretized version of the quantum mechanical path integral, and then
resorting to hamiltonian path integral methods, we succeeded in
filling in missing steps from previous discussions.
Along the way, we uncovered a possible connection between
coordinate transformations and renormalization group transformations,
and we are currently investigating the possible interpretation that the
extra terms may emerge from a ``blocking'' procedure~\cite{us}.
A main virtue of our path integral based proof is that it provides for
an extension to quantum field theory.
As a precursor to carrying out the discretization in the quantum field
theory case, we indirectly investigated extra terms in quantum field
theory by examining what transpires
as one dimensionally reduces a $1+1$ dimensional field theory where
usually no extra terms are considered, to a quantum mechanical theory
where the existence of extra terms is well-established.
We performed field redefinitions both before and after reduction,
and were able to derive an explicit form of an extra term appearing in
the quantum field theory.
In a forth-coming article~\cite{us}, we will argue that
in spite of the path integral's ubiquitous employment,
manipulations involving the path integral must be reevaluated,
and that interesting physics could be missed.
Applications are numerous since nonlinear field redefinitions are
commonplace.  Exciting possibilities include higher order
${\cal O} (\hbar^n) $ $n \geq 2$ anomalies in theories with nonlinear
symmetries,
corrections to quantities computed using collective coordinate
techniques, and corrections to duality transformations (such as the
higher order dilaton corrections associated with the $R$ symmetry in
string theory).  These issues ares currently under investigation by
the authors~\cite{us}.

\section*{Acknowledgements}
\hspace{\parindent}%
K.A. dedicates this paper to her beloved father Dr.~Max Apfeldorf,
who died January 18, 1995, and
C.O. dedicates this paper to his brother Jorge Ricardo Ord\'o\~nez
Montenegro who died March 13, 1995.
We thank Steven Weinberg and Jacques Distler for useful
discussions of the integrating out procedure and the renormalization group.
This research was supported in part by Robert A.~Welch Foundation,
NSF Grant PHY 9009850 (UT Austin) and
Department of Education grant DE-FG05-87ER40367 (Vanderbilt).

\bigskip
\appendix
\section{Useful identities involving modes}
In this appendix, we prove some identities used in obtaining
results in this paper.
Consider the expansion in modes of a function $\varphi (\theta , t)$
with a periodic coordinate $\theta$
\beq
\varphi(\theta,t) = \sum_{p \in Z} b^{(p)}(t) e^{i p \theta} .
\eeq
We call $p$ the mode number.
Define the following object which is multilinear in $N$ modes
and where the sum of the mode numbers is $\ell$
\beq
E^{(\ell)}_N [b]  \equiv  \sum_{p_1 \in Z} \ldots \sum_{p_N \in Z}
b^{(p_1)} \ldots b^{(p_N)} \delta^{\ell - \sum_{i=1}^N p_i =0} .
\eeq
Note the special case when the sum of the mode numbers is zero
\beq
E^{(0)}_N [ b ] = \int \frac{d\theta}{2 \pi} \varphi(\theta,t)^N
\equiv B_N[b] .
\label{eq:Bmode}
\eeq
One may straightforwardly relate the sum over a product of two $E$'s
to a single $E$ in the following way
\beqar
& & \sum_{p \in Z} E^{(\ell + p)}_M [b] E^{(-p+m)}_N [b]  \nonumber \\
&=&  \sum_{p \in Z}
\sum_{\begin{array}{c} {p_1,\ldots p_M \in Z} \\
\ell_1,\ldots \ell_N \in Z  \end{array}}
b^{(p_1)} \ldots b^{(p_N)} b^{(\ell_1)} \ldots b^{(\ell_N)}
\delta^{p+ \ell - \sum_{i=1}^M p_i =0} \delta^{-p +m - \sum_{j=1}^N
\ell_j = 0}  \nonumber \\
&=& \sum_{p \in Z} \sum_{p_1 \in Z} \ldots \sum_{p_{M+N} \in Z}
b^{(p_1)} \ldots b^{(p_{M+N})}
\delta^{p+ \ell - \sum_{i=1}^M p_i =0} \delta^{-p +m - \sum_{j=M+1}^{M+N}
p_j = 0} \nonumber \\
&=& \sum_{p_1 \in Z} \ldots \sum_{p_{M+N} \in Z}
b^{(p_1)} \ldots b^{(p_{M+N})} \delta^{\ell + m - \sum_{j=1}^{M+N} p_j = 0}
\nonumber \\
&=&  E^{(\ell + m)}_{M+N} [b] .
\eeqar
Consider three $E$'s where the number of modes may be different
and where the sum of all mode numbers is zero
\beqar
& & \sum_{\ell \in Z} \sum_{n \in Z} \sum_{p \in Z}
E^{(\ell + n)}_L [b] \: \: \: E^{(-\ell+p)}_M [b] E^{(-p-n)}_N [b]
\nonumber  \\
&=& \sum_{n \in Z} \sum_{p \in Z} E^{(n+p)}_{L+M} [b] \:\:\: E^{(-p-n)}_N [b]
\nonumber \\
&=& \left[ \sum_{n \in Z} 1\right]
\sum_{p^\prime \in Z} E^{(p^\prime)}_{L+M} [b] \:\:\: E^{(-p^\prime)}_N [b]
\nonumber  \\
&=& \left[ \sum_{n \in Z} 1 \right] B_{L+M+N} [b] .
\eeqar
Finally consider the following trace of $j$ $E_M$'s where
the sum of all mode numbers is zero (summation convention assumed)
\beqar
{\rm tr } (E_M^j) &=& E^{(-n_j + n_1)}_M [b] E^{(-n_1 + n_2)}_M [b] \ldots
E^{(-n_{j-1}+n_j)}_M [b] \nonumber  \\
&=& E^{(-n_j + n_2)}_{2M} [b] E^{(-n_2 + n_3)}_M [b] \ldots
E^{(-n_{j-1}+n_j)}_M [b]  \nonumber \\
&=& \left[ \sum_{n \in Z} 1 \right] B_{j(N-1)} [b] .
\eeqar
This identity is useful when evaluating the jacobian determinant in
section 3.1 of this paper.

A final identity which is useful in section 3.3 where we
compare the kinetic terms in the two methods is
\beqar
& & \sum_{p \in Z} p^2 E_N^{(p)} E_1^{(-p)} \nonumber \\
&= &  \sum_{p \in Z} \sum_{\ell_1 \in Z} \cdots \sum_{\ell_N \in Z}
p^2  b^{(\ell_1)} \cdots b^{(\ell_N)} b^{(-p)}
\delta^{p-\sum_{i=1}^N \ell_i =0}  \nonumber \\
&= &  \sum_{\ell_1 \in Z} \cdots \sum_{\ell_N \in Z}
\left( \sum_{i=1}^N \ell_i \right)^2
b^{(\ell_1)} \cdots b^{(\ell_N)} b^{(-\sum_{i=1}^N \ell_i)}
\nonumber \\
&=& \sum_{\ell_1 \in Z} \cdots \sum_{\ell_N \in Z}
\left( \sum_{i=1}^N (\ell_i)^2  + \sum_{i \neq j} \ell_i \ell_j \right)
b^{(\ell_1)} \cdots b^{(\ell_N)} b^{(-\sum_{i=1}^N \ell_i)}
\nonumber \\
&=& \sum_{\ell_1 \in Z} \cdots \sum_{\ell_N \in Z}
\left( N (\ell_N)^2  + N \ell_N \sum_{i=1}^{N-1} \ell_i  \right)
b^{(\ell_1)} \cdots b^{(\ell_N)} b^{(-\sum_{i=1}^N \ell_i)}
\nonumber \\
&=& N \sum_{\ell_1 \in Z} \cdots \sum_{\ell_N \in Z}
\ell_N \left( \sum_{i=1}^{N-1} \ell_i  \right)
b^{(\ell_1)} \cdots b^{(\ell_N)} b^{(-\sum_{i=1}^N \ell_i)}
\nonumber \\
&=& -N \sum_{p \in Z} \sum_{\ell_1 \in Z} \cdots \sum_{\ell_N \in Z}
\ell_N p b^{(\ell_1)} \cdots b^{(\ell_N)} b^{(p)}
\delta^{(-p-\sum_{i=1}^N \ell_i =0 )}
\nonumber \\
&=& -N \sum_{p \in Z} \sum_{\ell_1 \in Z} \cdots \sum_{\ell_{N-1} \in Z}
\sum_{\ell \in Z} \ell p b^{(\ell_1)} \cdots b^{(\ell_{N-1})} b^{(\ell)}
b^{(p)} \delta^{(-p-\ell -\sum_{i=1}^{N-1} \ell_i =0 )} \nonumber \\
&=& -N \sum_{p \in Z} \sum_{\ell \in Z} \ell p
E_1^{(p)} E_1^{(\ell)} E_{N-1}^{(-\ell-p)}
\label{eq:p2ident} .
\eeqar

\section{Details of Kaluza Klein calculation}
The metric in terms of the $b^{(m)}$ modes in the Kaluza Klein calculation
is
\beqarn
g_{mn}[b] &=& 2 \pi R
(\delta_{m+n=0} + 2 \alpha N E^{(-m-n)}_{N-1}[b] + N^2 \alpha^2
E^{(-m-n)}_{2N-2}[b] ) \\
&=& 2 \pi R (1 +  \alpha N E_{N-1}[b] )^2_{m+n} ,
\eeqarn
where $1_{m+n} $ is shorthand for $\delta_{m+n=0},$
$(E_M)_{m+n}$ is shorthand for $E_M^{(-m-n)}$, and
$(E_M^2)_{m+n}$ is shorthand for $E_M^{(-m-p)} E_M^{(p-n)}$.
Subtracting out an infinite constant due to the $2 \pi R,$
we compute the jacobian contribution
\beqar
{\rm Tr ln } [\frac{g_{mn}[b]}{2 \pi R}] \delta(t-t^\prime)
&=& 2 {\rm Tr ln } (1 +  \alpha N E_{N-1}[b] )_{m+n}  \delta(t-t^\prime)
\nonumber \\
&=& -2 {\rm Tr }\sum_{j=1}^\infty\frac{(-\alpha N)^j}{j} (E^j_{N-1}[b])_{m+n}
\delta(t-t^\prime) \nonumber \\
&=& -2 \delta(0) \sum_{j=1}^\infty \frac{(-\alpha N)^j}{j} \int dt
\left[  E^{(-m+n_1)}_{N-1}[b] \cdots   E^{(n_{j-1}+m)}_{N-1}[b]   \right]
\nonumber \\
&=& -2 \delta(0) \sum_{j=1}^\infty \frac{(-\alpha N)^j}{j}
\left[ \sum_{k \in Z} 1 \right] \int dt \: B_{j(N-1)}[b]   .
\eeqar
We now illustrate how the inverse metric is obtained to all orders in
$\alpha$ via a recursion equation.  Write
\beq
g^{km}[b] = \frac{1}{2 \pi R}
\sum_{Q=0}^\infty f_Q (\alpha) E^{(k+m)}_{Q}[b].
\eeq
and compute $g^{km}[b]  g_{mn}[b] = \delta^{k-n=0}$ as follows
\beqarn
\delta^{k-n=0} &=& \sum_{Q=0}^\infty f_Q (\alpha) E^{(k-n)}_{Q}[b]
+ 2 \alpha N \sum_{Q=0}^\infty f_Q (\alpha) E^{(k+m)}_{Q}[b]
E^{(-m-n)}_{N-1}[b] \\
& & +  \alpha^2 N^2 \sum_{Q=0}^\infty
f_Q (\alpha) E^{(k+m)}_{Q}[b] E^{(-m-n)}_{N-2}[b] \\
&=& \sum_{Q=0}^\infty f_Q (\alpha) E^{(k-n)}_{Q}[b]
+ 2 \alpha N \sum_{Q=0}^\infty f_Q (\alpha) E^{(k-n)}_{Q+N-1}[b]  \\
& & +  \alpha^2 N^2 \sum_{Q=0}^\infty f_Q (\alpha) E^{(k-n)}_{Q+2N-2}[b] \\
&=& \sum_{Q=0}^{N-2} f_Q (\alpha) E^{(k-n)}_{Q}[b]
+ \sum_{Q=0}^{N-2} \left( f_{Q+N-1} (\alpha)
+ 2 \alpha N f_{Q} (\alpha) \right)  E^{(k-n)}_{Q+N-1}[b]  \\
& & + \sum_{Q=0}^\infty \left( f_{Q+2N-2} (\alpha) +
2 \alpha N f_{Q+N-1} (\alpha) +
\alpha^2 N^2 f_{Q} (\alpha) \right)  E^{(k-n)}_{Q+2N-2}[b].
\eeqarn
The first sum on the righthand side after the last equal sign
must be a Kroeneker delta function, and the other two sums
must vanish.  From the first sum, one finds $f_0(\alpha) =1$ and
$f_1(\alpha) = \cdots = f_{N-2}(\alpha) =0.$   From the second sum,
one finds $f_{N-1}(\alpha) =-2 \alpha N$ and
$f_{N}(\alpha) = \cdots = f_{2N-3}(\alpha) =0.$   From the third sum,
it is clear that $f_{Q}(\alpha) =0$
if $Q$ is not divisible by $N-1.$
Furthermore, making the ansatz
$$
f_{(N-1)q}(\alpha) = (-\alpha N)^q C_q ,
$$
we may determine the unknown function $C_q$ by plugging into
the third sum.
Explicitly, the condition that the third sum vanishes is
$$
f_{(N-1)(2+q)}(\alpha) = -2 \alpha N f_{(N-1)(1+q)}(\alpha)
- \alpha^2 N^2 f_{(N-1)q}(\alpha),
$$
and taking into account the boundary condition $C_0=1,$
this leads to the solution
$$
f_{(N-1)q}(\alpha) = (-\alpha N)^q (q+1) .
$$
Thus, we obtain the inverse metric to all orders in the parameter $\alpha$
\beq
g^{km}[b] = \frac{1}{2 \pi R}
\sum_{j=0}^\infty (-\alpha N)^j (j+1) E^{(k+m)}_{j(N-1)}[b].
\eeq
Using this expression, the connection coefficients are computed to be
\beq
\Gamma^k_{\ell p}[b] = - \sum_{j=0}^\infty (-\alpha N)^{j+1} (N-1)
E^{(k-p-\ell)}_{j(N-1)+N-2}[b].
\eeq
Note that $\Gamma^k_{\ell p}[b]$ is dependent only on the total mode
number $k-p-\ell$ so it will prove useful to write
$\Gamma^k_{\ell p}[b] = \Gamma^{k-p-\ell}[b]$.
The metric and inverse metric also depend only on the sum of their indices.
Using these expressions, the extra term may be found by computing
\beqar
& & \sum_{k,\ell, p, n \in Z}
\frac{1}{8} g^{\ell n}[b] \Gamma^k_{\ell p}[b] \Gamma^p_{nk}[b]
\nonumber \\
&=& \sum_{k,\ell, p, n \in Z} \frac{1}{8} g^{\ell n}[b] \Gamma^{k -\ell- p}[b]
\Gamma^{p -n -k}[b] \nonumber \\
&=& \sum_{k \in Z} 1 \sum_{\ell, p, n \in Z} \frac{1}{8} g^{\ell n}[b]
\Gamma^{-\ell-p}[b] \Gamma^{p-n}[b]  \nonumber \\
&=& \left[ \sum_{k \in Z} 1 \right]^2
\frac{(N-1)^2 }{8\cdot 2 \pi R} \sum_{L,M,J=0}^\infty
(-\alpha N)^{L+M+J+2} (L+1)  E^{(0)}_{(M+J+L)(N-1) +2(N-2) }[b]
\nonumber \\
&=& \left[\sum_{k \in Z} 1 \right]^2 \frac{(N-1)^2 }{8\cdot 2 \pi R}
\sum_{L,M,J=0}^\infty (-\alpha N)^{L+M+J+2}
\left(\frac{L+M+J}{3} +1 \right) \nonumber \\
& &\:\quad \quad \quad \quad \quad  \quad  \quad  \quad  \quad  \quad
\quad  \quad
\times B_{(M+J+L)(N-1) +2(N-2) }[b]  \nonumber \\
&=& \left[ \sum_{k \in Z} 1 \right]^2 \frac{(N-1)^2}{8\cdot 2 \pi R}
\sum_{X=0}^\infty {\cal P}_3(X) (-\alpha N)^{X+2} \left(\frac{X}{3}+1 \right)
B_{X(N-1) +2(N-2) }[b] \nonumber \\
&=& \frac{(N-1)^2}{8\cdot 2 \pi R} \left[ \sum_{k \in Z} 1 \right]^2
\sum_{X=0}^\infty  (-\alpha N)^{X+2}
\left( \begin{array}{c} X+3 \\ 3 \end{array} \right) B_{X(N-1) +2(N-2) }[b] .
\eeqar
We have used the fact that ${\cal P}_3(X) = \half (X+1)(X+2) $
is the number of ways to write $X$ as the sum of 3 non-negative integers.
To rewrite this expression in terms of $\varphi$, we use
equation~\ref{eq:Bmode}
\beqar
&=& \frac{(N-1)^2}{48 \cdot 2 \pi R}
\left[ \sum_{k \in Z} 1 \right]^2
\sum_{X=0}^\infty  (-\alpha N)^{X+2} (X+1) (X+2) (X+3)
\nonumber \\
& & \times \int \frac{d\theta}{2 \pi} \varphi^{X(N-1) +2(N-2) }[b] \nonumber \\
&=& \frac{(N-1)^2 }{48 \cdot 2 \pi R} \left[ \sum_{k \in Z} 1 \right]^2
\int \frac{d\theta}{2 \pi}
\frac{1}{(-\alpha N)} \varphi^{2N-4} \frac{d^3 \:\:\:}{d(\varphi^{N-1})^3}
\sum_{X=0}^\infty  (- \alpha N \varphi^{N-1} )^{X+3}  \nonumber \\
&=& \frac{(N-1)^2}{48\cdot 2 \pi R} \left[ \sum_{k \in Z} 1 \right]^2
\int \frac{d\theta}{2 \pi}  \:\: \frac{1}{(-\alpha N)}
\varphi^{2N-4} \frac{d^3 \:\:\:}{d(\varphi^{N-1})^3}
\left[ \frac{ (-\alpha N \varphi^{N-1})^3}{1+ \alpha N \varphi^{N-1}}
\right]  \nonumber \\
&=& \frac{1}{8} \frac{\alpha^2 N^2 (N-1)^2}{2 \pi R}
\left[ \sum_{k \in Z} 1 \right]^2
\int \frac{d\theta}{2 \pi}  \:\:
\frac{(\varphi^{N-2} )^2}{(1 + \alpha N \varphi^{N-1})^4} .
\eeqar

\end{document}